\documentclass[aps,preprint,showpacs,amsmath,amssymb]{revtex4}
\usepackage{epsfig}
\usepackage{wasysym}
\usepackage{amssymb}

\begin{document}

\newcommand{\bvec}[1]{\mbox{\boldmath ${#1}$}}
\title{Nonidentical protons}
\author{T. Mart and A. Sulaksono}
\affiliation{Departemen Fisika, FMIPA, Universitas Indonesia, Depok 16424, 
  Indonesia}
\date{\today}
\begin{abstract}
  We have calculated the proton charge radius by assuming that 
  the real proton radius is not unique and the radii are randomly distributed 
  in a certain range. This is
  performed by averaging the elastic electron-proton differential
  cross section over the form factor cut-off. By using a dipole form factor and
  fitting the middle value of the cut-off to the low $Q^2$  
  Mainz data, we found the lowest $\chi^2/N$ for a cut-off 
  $\Lambda=0.8203\pm 0.0003$ GeV, which corresponds to a proton 
  charge radius $r_E=0.8333\pm 0.0004$ fm. 
  The result is compatible with the recent
  precision measurement of the Lamb shift in muonic hydrogen 
  as well as recent calculations using more sophisticated techniques.
  Our result indicates that the relative variation of the form factor
  cut-off should be around 21.5\%. Based on this result we have investigated 
  effects of the nucleon radius variation on the 
  symmetric nuclear matter (SNM) and the neutron star matter (NSM)
  by considering the excluded volume effect in our calculation.
  The mass-radius relation of neutron star is found to be 
  sensitive to this variation. The nucleon effective mass
  in the SNM as well as the equation of state of both the SNM and the NSM
  exhibit a similar sensitivity.
\end{abstract}

\maketitle

\section{Introduction}
The recent precise measurement of the Lamb shift in muonic hydrogen 
atom \cite{pohl}
has sparked a controversy, because this measurement yields
a smaller proton charge radius, i.e. $r_E=0.84184(67)$ fm. This
radius is significantly smaller than the standard CODATA value
\cite{codata}, 
$r_E=0.8768(69)$ fm, which is based on the measurements of the 
Lamb shift in electronic 
(conventional) hydrogen atom, as well as the results from 
elastic electron-proton scatterings. The latest precise measurement 
of elastic electron-proton
scattering at MAMI, Mainz, which yields
$r_E=0.879(5)_{\rm stat.}(4)_{\rm syst.}
(2)_{\rm model}(4)_{\rm group}$ fm \cite{bernauer}, clearly
supports the CODATA value. Considerable efforts 
\cite{miller_pra,miller:1207.0549,jaeckel,derujula,cloet,barger_106,tucker,batell,hill,carroll,%
rivas,pineda,Jentschura,brax,Eides:2012ue,carlson_pra,barger_108,borie} 
have been devoted to attack this proton radius problem.  
References~\cite{miller_pra,miller:1207.0549}, for instance, 
propose that the off-shell
form factors of the proton could generate large polarizability contributions 
to the proton structure and eventually could solve the problem, 
since the effect 
would only appear in the case of muonic Hydrogen. However, 
a different opinion has been put forward in Ref. \cite{birse}, in 
which the off-shell effect is shown to be not sufficiently large to
reduce the discrepancy between the radii found in muonic and
conventional Hydrogen atoms. 
It is interesting to note that Ref. \cite{birse} also concludes
that the resolution of this problem must lie elsewhere, perhaps in 
re-analyses of the older experiments. Furthermore, 
QED is believed to be more precise than QCD and 
the techniques and methods of the Lamb shift  measurement 
in muonic hydrogen as well as electron-proton
scattering are beyond any doubts \cite{Distler:2010zq}. Therefore, 
it is urgent to reinvestigate the prevailing
 methods of 
extracting the proton charge radius. Such an idea has been
recently proposed by a number of research groups 
\cite{Lorenz:2012tm,Kelkar:2012hf}.
Nevertheless, surprisingly none of them has questioned the
idea of the ''radius'' itself.

The radius of proton is defined in accordance with our 
imagination that
proton has a spherical form. However, recent investigations
have revealed that protons could deform from a spherical 
shape, like nuclei in nuclear physics. This originates 
from the relativistic motion of the spin 1/2 quarks inside
the proton \cite{miller}, although there is also a claim
that a pure $s$-wave nucleon model, with $l=0$ and thus perfectly
spherical, could be constructed \cite{gross}. 
Meanwhile, in
the liquid drop model it is also customary to assume
the variation of the proton and neutron radii in order
to explain, e.g., the polarized electric dipole moment
in the reflection asymmetric nuclei \cite{denisov}.
Obviously, the definition of radius gets blurred
if the proton were not spherical.

On the other hand, the fluctuating size of proton has become
an important idea in explaining the oscillating color 
transparency \cite{ralston}. The idea behind this fluctuating
size is that in the proton-nucleus scattering the high energy
protons that scattered at wide angles should be ''small''. 
However, there is also a certain process, in which the protons
must be ''large'' and the amplitude of this process
will increase with increasing the proton sizes.  Using this idea
the oscillating transparency found in experiment \cite{caroll}, 
which is defined as the ratio
between the proton-proton scattering 
cross sections off the nucleus and off the proton
at $90^\circ$ as a function of energy, can
be successfully reproduced  \cite{ralston}.

Based on the above experiences in this paper we propose a
calculation of the proton charge radius by assuming that 
protons do not have identical radii, they vary in a certain 
range. To simplify the problem we further assume that the
radii are randomly distributed around their average value. 
Practically, since the proton radius enters 
the cross section via the charge form factor, we can perform
this calculation by taking the form factor cut-off as the
corresponding variable. We believe that further corrections
could enter the cross section formulation. However, for the
present exploratory study we also believe that our assumption would be
sufficient.

We note that our result is in agreement with that obtained from muonic
hydrogen \cite{pohl}. A more careful measurement has been carried
out at Paul Scherer Institute and the result has just been published \cite{aa}.
It is interesting to note that the latter is still consistent with the
previous measurement  \cite{pohl}, indicating that the proton radius
extracted from muonic hydrogen would hardly change. Therefore, the
discrepancy between our result and the electronic hydrogen
experiment is still outstanding and more efforts are required
to alleviate this problem.

Whereas a five-percent difference in the proton radius could trigger
a strong controversy in hadronic studies, it is quite 
ironic to realize that in the nuclear and neutron star matter 
investigations protons and neutrons are traditionally considered as ''point
particles''. Effects of the nucleon structures are only considered
in the so-called {\it excluded volume effect} (EVE) model
\cite{bxsun,aguirre,aguirre1,panda,patra,anghel,kouno}. In this model
the total volume occupied by $N$ nucleons, i.e. $Nv_N$, is subtracted
from the nuclear matter volume $V$, so that the effective volume  
available for the nucleon  motion is reduced to $V-Nv_N$,
where $v_N$ is the volume of a nucleon. Of course, the volume
of the nucleon itself decreases as the matter density increases. 
However, surprisingly there has been 
no unique definition of the radius in free space, i.e. 
at zero-density. For instance, Ref.~\cite{aguirre} used the proton radii
$r_p=0.80$ fm, 0.70 fm, and 0.60 fm to study the EVE 
on the equation of state of homogeneous hadronic matter, whereas 
Ref.~\cite{bxsun} used $r_p=0.63$ fm to study the effect on 
the equation of state of nuclear matter. Nevertheless, all
studies indicate that the effect is non-negligible. In fact, 
Ref.~\cite{aguirre} found that the effect can enlarge the range
of applicability of the quark-meson-coupling model.
Thus, it would be very interesting to study the 
EVE by using our knowledge obtained from the elastic 
electron-proton scattering process.

In Sec.~\ref{sec:extraction} of this paper we explain the procedure of
extracting the proton charge form factor. Section \ref{sec:future}
briefly discuss the possible future experiment for refining
the present calculation. In Sec.~\ref{sec:nuclear} we investigate
effects of the nucleon radius variation in the 
neutron star and symmetric nuclear matter. 
We will summarize and conclude our findings in
Sec.~\ref{sec:conclusion}.

\section{Extraction of the proton radius from electron-proton scattering}
\label{sec:extraction}

The differential cross section for elastic electron-proton scattering 
can be efficiently written in terms of the Sachs electric and magnetic form
factors, $G_{E,p}$ and $G_{M,p}$, as \cite{halzen}
\begin{eqnarray}
  \label{eq:cross-section}
  \frac{d\sigma}{d\Omega} = \left(\frac{d\sigma}{d\Omega}\right)_{\rm Mott}
  \frac{1}{(1+\tau)} \left[G_{E,p}^2(Q^2)+ \frac{\tau}{\epsilon} G_{M,p}^2(Q^2) 
  \right] ,
\end{eqnarray}
where the Mott cross section, 
\begin{eqnarray}
  \label{eq:mott}
  \left(\frac{d\sigma}{d\Omega}\right)_{\rm Mott} = \frac{E'}{E}
  \frac{\alpha^2}{4E^2}\frac{\cos^2(\theta/2)}{\sin^4(\theta/2)} ~,
\end{eqnarray}
describes the elastic scattering of point-like particle. The notation 
$\epsilon=[1+2(1+\tau)\tan^2(\theta/2)]^{-1}$ denotes the virtual photon 
polarization, $Q^2=4EE'\sin^2(\theta/2)$ is the square of the virtual photon 
momentum transfer, $\tau=Q^2/4m_p^2$, $m_p$ is the proton mass, 
and $E$ ($E'$) represents the electron initial (final) lab energy
with scattering angle $\theta$.

Since we will not focus on the problems of extracting the magnetic
form factor, we will use the phenomenological scaling 
$G_{M,p}=\mu_pG_{E,p}$, where $\mu_p$ is the proton anomalous magnetic
moment, to simplify Eq.~(\ref{eq:cross-section}) to
\begin{eqnarray}
  \label{eq:cross-section-recast}
  \frac{d\sigma}{d\Omega} = \left(\frac{d\sigma}{d\Omega}\right)_{\rm Mott}
  \frac{1}{(1+\tau)} \left[ 1+\frac{\tau}{\epsilon}\mu_p^2\right] 
  G_{E,p}^2(Q^2,\Lambda) ~,
\end{eqnarray}
assuming identical charge distributions in the protons, i.e. 
identical electric and magnetic radii, where $\Lambda$ 
is the corresponding cut-off.

However, if the the proton sizes were not identical, and if we
assume that they were randomly distributed near 
their middle value, then 
Eq.~(\ref{eq:cross-section}) must be averaged over all proton sizes, 
i.e., averaged over the form factor cut-off $\Lambda$, 
\begin{equation}
  \label{eq:cross-section-av}
  \left\langle\frac{d\sigma}{d\Omega} \right\rangle
  = \left(\frac{d\sigma}{d\Omega}\right)_{\rm Mott}
  \frac{1}{(1+\tau)} \left[ 1+\frac{\tau}{\epsilon}\mu_p^2\right] 
  \langle G_{E,p}^2(Q^2,\Lambda_1)\rangle ,
\end{equation}
where
\begin{eqnarray}
  \label{eq:dipole-av}
  \langle G_{E,p}^2(Q^2,\Lambda_1)\rangle = \frac{1}{2\Delta\Lambda}
  \int_{\Lambda_1-\Delta\Lambda}^{\Lambda_1+\Delta\Lambda}
  G_{E,p}^2(Q^2,\Lambda) d\Lambda
  ~,
\end{eqnarray}
and $2\Delta\Lambda$ represents the range of the cut-off variation
around the middle value $\Lambda_1$.

Experimental measurements for decades have indicated
that the square root of this average can be parameterized by means
of a dipole form,
\begin{eqnarray}
  \label{eq:dipole-av-equal}
  \langle G_{E,p}^2(Q^2,\Lambda_1)\rangle^{1/2} 
  \approx\left(1+\frac{Q^2}{\Lambda_1^2}\right)^{-2}
  ~,
\end{eqnarray}
where $\Lambda_1^2=0.71$ GeV$^2$ is often called as the standard dipole
form factor, 
from which one obtains the proton electric radius by calculating 
the form factor slope at the real photon point, 
\begin{eqnarray}
  \label{eq:rms}
  r_E \equiv \langle r_{E,p}^2 \rangle^{1/2} = 
  \left(\left.
    -6\frac{dG_{E,p}(Q^2)}{dQ^2}\right|_{Q^2=0}
  \right)^{1/2}~.
\end{eqnarray}
At this stage it is important to note that after the operation of modern
continuous beam accelerators, such as MAMI in Mainz and CEBAF at the 
Jefferson Laboratory,
significant deviation from the standard dipole form factor
has been observed. To account for this deviation a number of new fits
and models has been proposed. This includes modifications
of the dipole form \cite{friedrich,arrington-2004}, as well as
introduction of more physical ingredients in the form factor 
\cite{gari,arrington-2007}.
There seems to be no need to keep the original dipole form
to fit both electric and magnetic form factors, especially for
a global fit to all data, since the standard dipole is 
considered as just a phenomenological approximation. 
Furthermore, the choice of the dipole form 
also seems to be trivial.

However, in the non-relativistic limit as well as 
in the Breit frame the dipole form factor is related 
to an exponentially decaying charge distribution via
the Fourier transform. The exponentially decaying behavior is 
found in most natural phenomena, from radioactive decay rate 
to atmospheric pressure on earth. Some phenomena in social 
science also exhibit this behavior. 
Thus, we believe that in our present case 
a dipole form factor looks more
natural and a deviation from such a natural phenomenon 
requires a rigorous physical concept.

Furthermore, it is also important to emphasize here that 
the determination of the slope at $Q^2=0$ given by 
Eq.~(\ref{eq:rms}) requires very good knowledge of
$G_{E,p}(Q^2)$ at very low $Q^2$. As a consequence,
the closer we can approach the real photon point experimentally,
the more reliable we can determine the proton charge radius.
Therefore, the latest and accurate measurement of  the 
elastic electron-proton scattering with low energy and low $Q^2$
at MAMI \cite{bernauer} provides very suitable experimental 
data for our present discussion. 

We begin with Eq.~(\ref{eq:dipole-av-equal}), which implies that the
experimentally observed form factor is in fact an average 
to the genuine form factor,
that one actually should use in Eq.~(\ref{eq:rms}) in order 
to get the real proton radius. As a consequence, the extracted
proton radius in this way should be considered as an averaged radius.

To investigate the effect of averaging the form factor
given by Eq.~(\ref{eq:dipole-av}), let us use the standard 
dipole form factor to calculate
$\langle G_{E,p}^2(Q^2,\Lambda_1)\rangle$ in Eq.~(\ref{eq:dipole-av}). 
Note that if we
use a dipole form, the magnitude of the relative variations 
of both radius $r$ and cut-off $\Lambda$ are equal, i.e.
\begin{eqnarray}
  \label{eq:ratio}
  \left|\frac{\Delta \Lambda}{\Lambda}\right| =
  \left|\frac{\Delta r}{r}\right| ~,
\end{eqnarray}
provided that the variations are not extremely large. 

\begin{figure}[t]
  \begin{center}
    \leavevmode
    \epsfig{figure=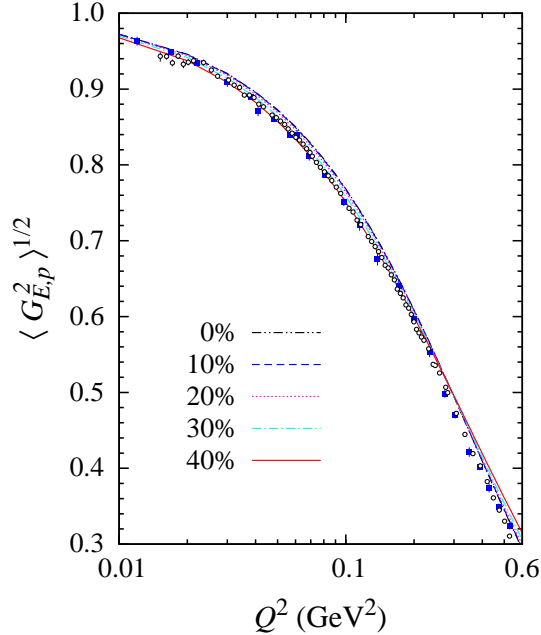,width=75mm}
    \caption{(Color online) Square root of the averaged $G_{E,p}^2$
    calculated from Eq.~(\ref{eq:dipole-av}) for different values
    of the relative variation $\Delta\Lambda/\Lambda_1$ 
    (shown in the figure) compared with
    experimental data. Experimental data are taken from 
      Refs.~\cite{bernauer} (open circles) and 
      \cite{arrington-2007} (solid squares).
    }
   \label{fig:dipole} 
  \end{center}
\end{figure}

The result is shown
in Fig.~\ref{fig:dipole}, where we compare our calculations 
obtained with the cut-off variations $\Delta\Lambda$ from zero up to
40\% of its standard value with experimental data.
Note that for the sake of simplicity we use the latest
result from Mainz experiment \cite{bernauer}, which provides the
latest and most accurate data in the low $Q^2$ region, and the result
of extraction from the world electron-proton scattering data 
with two-photon exchange effects included \cite{arrington-2007},
which represents the previous measurements.
We use the result obtained from the standard Rosenbluth separation 
technique for the Mainz data, in order to reduce the model dependency
of the data. The $G_{E,p}$ data extracted in Ref.~\cite{arrington-2007}
are of course model dependent. However, in this paper 
they are only used for the purpose 
of comparison and not included in the
fitting database as described in the following discussion.

\begin{figure}[t]
  \begin{center}
    \leavevmode
    \epsfig{figure=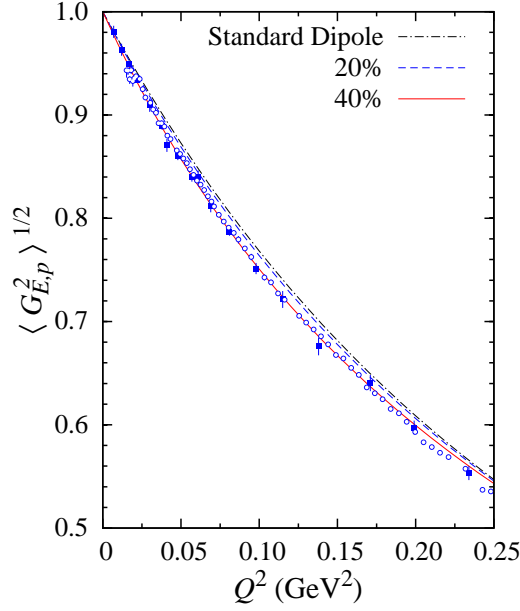,width=70mm}
    \caption{(Color online) As in Fig.~\ref{fig:dipole}, but 
      limited for 
      $Q^2\le 0.25$ GeV$^2$. The results obtained for two different values
      of $\Delta\Lambda/\Lambda_1$ are compared with the standard dipole
      form factor and experimental data. Note the linear scale for
    $Q^2$ axis.}
   \label{fig:dipole-low} 
  \end{center}
\end{figure}

It is obvious from  Fig.~\ref{fig:dipole} that
the result shows a variance to the standard dipole form.
For $Q^2\gtrsim 0.25$ GeV$^2$ we observe that the averaged 
form factors are larger than the standard dipole one 
(i.e. $\Delta\Lambda/\Lambda=0$); increasing the 
relative variation will increase the form factor.
However, for $Q^2\lesssim 0.25$ GeV$^2$ we observe
a different behavior, i.e. the form factor decreases as the 
variation increases. Since the decrease is 
relatively small, it is almost invisible in Fig.~\ref{fig:dipole}.
Therefore, in Fig.~\ref{fig:dipole-low} we increase the resolution
of the $\langle G_{E,p}^2(Q^2,\Lambda_1)\rangle^{1/2}$ axis 
by limiting 
$Q^2\lesssim 0.25$ GeV$^2$, where we can clearly see
the effect of variation, i.e. for the relative variation of $40\%$
the agreement with experimental data is almost perfect.

Although the agreement of the solid curve with experimental data in
Fig.~\ref{fig:dipole-low} could be fortuitous, the
most important message is that the standard dipole form factor is
still valid at low $Q^2$, provided that the corresponding
cut-off must be averaged with relative variation 
$\Delta\Lambda/\Lambda_1=40\%$. We believe that this is crucial 
because the extraction of proton charge radius is always plagued
with many complicated corrections, especially at 
high $Q^2$ regime, as discussed above. In view of this, in what follows,
we will only use the MAMI data and limit the $Q^2$ only 
up to 0.25 GeV$^2$.

\begin{figure}[h]
  \begin{center}
    \leavevmode
    \epsfig{figure=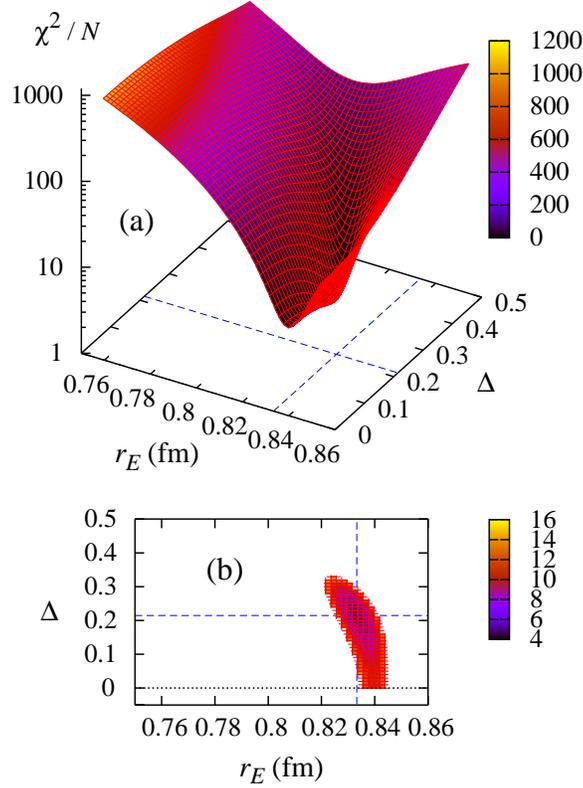,width=80mm}
    \caption{(Color online) (a) The obtained 
      $\chi^2/N$ as functions 
      of the proton charge radius $r$ and 
      $\Delta\equiv\Delta\Lambda/\Lambda_1$.
      (b) The projection of $\chi^2/N$
      on the $r-\Delta$ plane for $\chi^2/N\le 16$.
      Intersection of the vertical and horizontal 
      dashed lines  
      locates the minimum position of $\chi^2/N$,
      which is accurately shown in Fig.~\ref{fig:chi2}.}
   \label{fig:chi3d} 
  \end{center}
\end{figure}

It is apparent from Eq.~(\ref{eq:dipole-av}) and Fig.~\ref{fig:dipole-low} 
that for each value of $\Delta\Lambda$ we can optimize the
form factor cut-off $\Lambda_1$ in order to further 
improve the agreement of our calculation with 
experimental data. For this purpose we can calculate the standard
$\chi^2/N$ which measures the agreement of our calculation
with experimental data. The result as functions of the proton radius 
(translated from $\Lambda_1$) and $\Delta\Lambda/\Lambda_1$
is displayed in Fig.~\ref{fig:chi3d}. It is obvious that the
$\chi^2/N$ has only one minimum located by the 
intersection of the two dashed lines.

\begin{figure}[t]
  \begin{center}
    \leavevmode
    \epsfig{figure=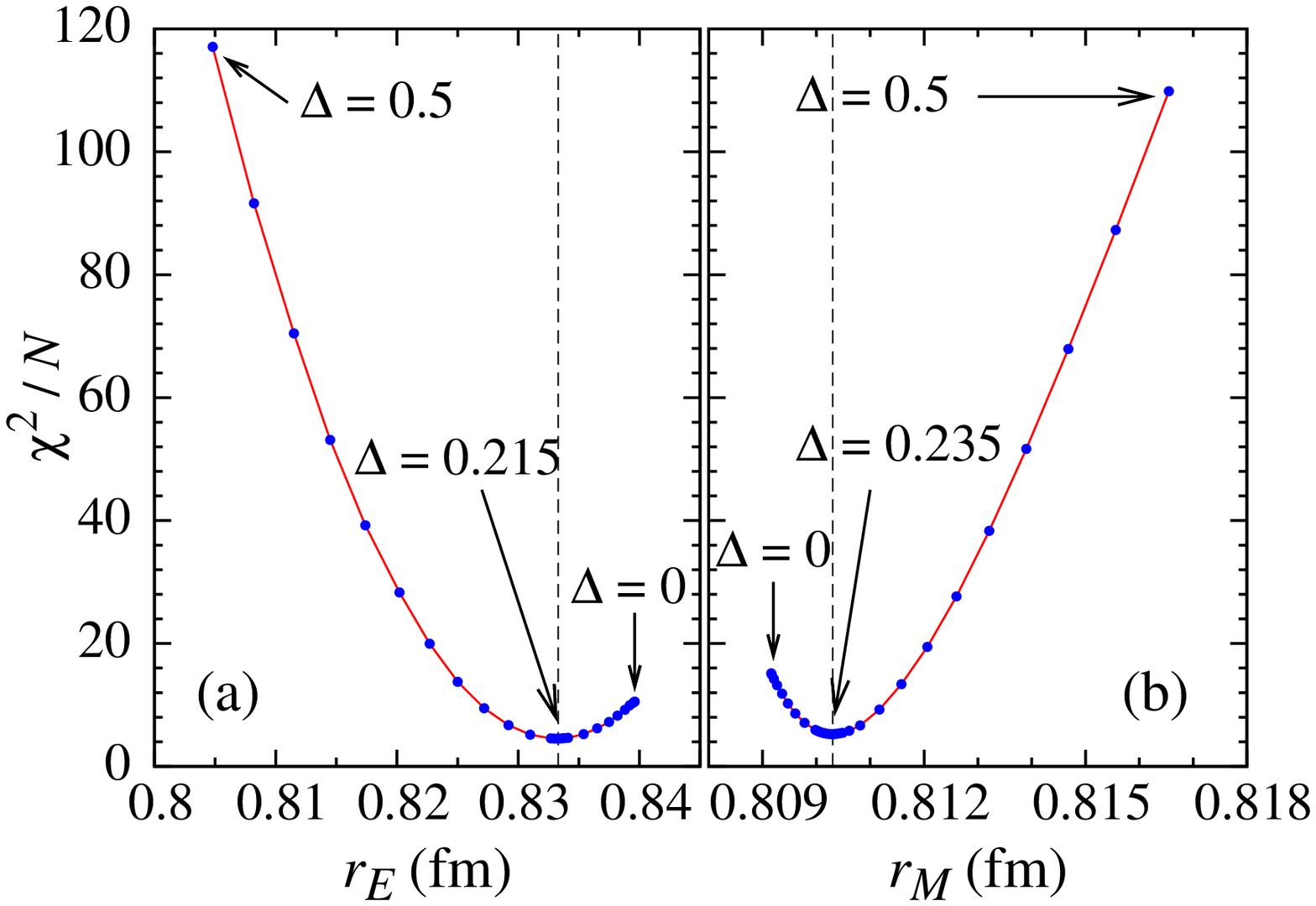,width=85mm}
    \caption{(Color online) The $\chi^2/N$ as a function
      of the obtained proton (a) charge and (b) magnetic
      radii. The lowest and highest values 
      of relative variation in the form factor cut-off, 
      $\Delta\equiv\Delta\Lambda/\Lambda_1$, are indicated in the figure.
      The vertical lines indicate the minimum positions of $\chi^2/N$.
      The corresponding $\Delta$ values are also shown.}
   \label{fig:chi2} 
  \end{center}
\end{figure}

To locate this minimum accurately we fit 
the value of $\Lambda_1$ by using the CERN-MINUIT
code and scan the relative variation $\Delta\Lambda/\Lambda_1$ 
from 0\% to 50\% with 2.5\% step, simultaneously, where the
value of $\Lambda_1$ is allowed to vary 
between 0.80 and 0.90. Although this choice
seems to be arbitrary, in our fits 
we found that the $\Lambda_1$ value
never reaches both upper and lower limits. 
For $\Delta\Lambda/\Lambda_1=0\%$ (50\%) the cut-off 
value is obtained to be 0.8142 (0.8493) GeV, which corresponds to the
proton charge radius of 0.8396 (0.8048) fm.

Having finished the scanning process we observe that the obtained 
$\chi^2/N$ forms a parabola with the minimum value at 
 $\Delta\Lambda/\Lambda_1=21.5\%$ and $\Lambda_1=0.8203$ GeV. This corresponds to the 
relative variation of proton radius $\Delta r_E/r_E\approx 21.5\%$, from
Eq.~(\ref{eq:ratio}), and the proton radius of $r= 0.8333$ fm,
from Eq.~(\ref{eq:rms}). The complete result 
of this scanning process is depicted in 
the left panel of Fig.~\ref{fig:chi2}.
To increase the accuracy, we have refined
the $\Delta\Lambda/\Lambda_1$ step to 0.005, which is equivalent
to $\Delta r_E=\pm 0.0003$ fm, in the vicinity of the minimum.
Therefore, our calculation would produce the best agreement with
experimental data if we used $r_E=0.8333\pm 0.0004$ fm, where we have
added the error bar coming from the fitting process 
($\Delta r$ obtained from the MINUIT
package). The present result is very interesting because
it corroborates most of the latest findings that exploits 
more sophisticated techniques \cite{Lorenz:2012tm,Kelkar:2012hf}.

We have also performed the above procedure to find the proton 
magnetic radius. It is well known that the experimental data
in this case are notoriously inaccurate, especially at 
$Q^2\approx 0$. As a consequence, we did not use the 
four lowest $Q^2$ data points from Ref.~\cite{bernauer-phd}, because most 
of them cannot be 
renormalized to $\mu_p$ at the real photon point.
For comparison with the charge radius, we display the result 
in the right panel of Fig.~\ref{fig:chi2}. 
Obviously, the trend is different as in the case of the
charge radius. In the case of the charge form factor,
increasing the relative variation is required to decrease
the magnitude of $G_{E,p}$ in order to reproduce experimental
data (see Fig.~\ref{fig:dipole-low}). In contrast to this, 
the relative variation of the radius is required
to increase the magnitude of $G_{M,p}$, in order
to reproduce the data.

As shown in  Fig.~\ref{fig:chi2} we obtain 
$r_M=0.8103\pm 0.0004$ fm, which is smaller than 
the result extracted from the dispersion relation, 
i.e. $0.84^{+0.01}_{-0.02}$ fm \cite{Lorenz:2012tm}. 
Nevertheless, our magnetic radius is much
larger than that obtained from direct extraction of
the Mainz data, i.e. $0.777\pm 0.013$ fm \cite{bernauer}.
However, if the Friedrich-Walcher parameterization \cite{friedrich}
was used in the latter, the magnetic radius would increase to 
$0.807\pm 0.02$ fm \cite{bernauer-phd}, which is apparently 
in good agreement with our finding. We believe that the 
less accurate magnetic form factor extracted from the
electron-proton scattering could be the origin of 
the large variance in the extracted magnetic radii
found in the literature.

\section{Expected future electron-proton scattering 
  experiments}
\label{sec:future}

\begin{figure}[t]
  \begin{center}
    \leavevmode
    \epsfig{figure=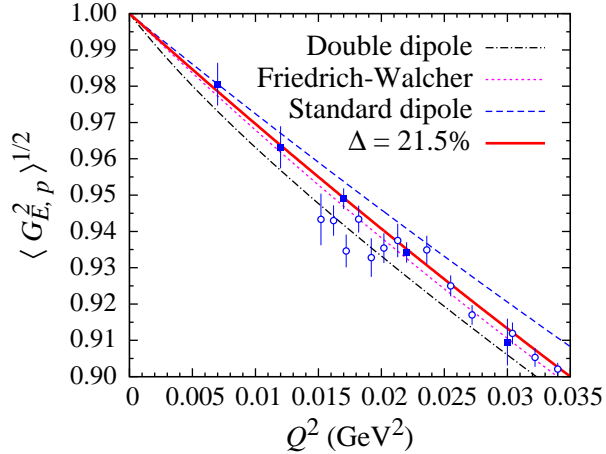,width=80mm}
    \caption{(Color online) Square root of the averaged $G_{E,p}^2$
      obtained from different calculations compared with
      experimental data for very low $Q^2$. 
      Notation for experimental data is as
      in Fig.~\ref{fig:dipole}.
      Result obtained in the present work is given by the solid
      curve.
      Parameters for the double-dipole and Friedrich-Walcher
      form factors are taken from Ref.~\cite{bernauer-phd}.}
   \label{fig:dipole_vlow} 
  \end{center}
\end{figure}

Since the mathematical formula of the proton charge form factor
is in principle not known, determination of the proton radius 
using Eq.~(\ref{eq:rms}) requires very good knowledge of the 
proton form factor to a very low $Q^2$ region. 
Thus, the real challenge for future experiments is to 
extend the current experimental data to this kinematics.
The situation is exhibited in Fig.~\ref{fig:dipole_vlow}, where
we compare the result for the 21.5\% proton radius variation
obtained in the previous section 
and various available parameterizations with experimental data
\cite{bernauer,arrington-2007}. It is obvious from this figure
that the Mainz data tend to deviate from our present result,
whereas, surprisingly, the data extracted in Ref.~\cite{arrington-2007}
show a very good agreement with our calculation. We note that
Ref.~ \cite{arrington-2007} used polynomial expansion to parameterize
the form factor during the extraction. Therefore, we believe that
the agreement with the present calculation as 
exhibited in Fig.~\ref{fig:dipole_vlow} could not
be a coincidence. 

As in the large $Q^2$ case, in the very low $Q^2$ region 
it is also obvious that the standard dipole form factor
is substantially larger than our calculation. The result of 
our calculation is very close to the result obtained from 
the Friedrich-Walcher model \cite{friedrich}. Since all four models shown
in  Fig.~\ref{fig:dipole_vlow} yield significantly different proton charge
radii, it is of course important to refine the experimental measurement
at this kinematics. Theoretically, this is possible because 
Eq.~(\ref{eq:cross-section}) explicitly shows that at low
$Q^2$ contribution of $G_{E,p}$ is dominant. Thus, at low $Q^2$ 
measurement of $G_{E,p}$ should be more accurate than that
of $G_{M,p}$.

However, from the experimental side this could be a daunting task.
It should be remembered that measurements of electron-proton
scattering cross section for $Q^2\approx 0.05$ GeV$^2$ were
already at forward angles \cite{bernauer-phd}. Below this point,
presumably one has to use other methods. One possible choice proposed
at MAMI is the use of the initial state radiation, i.e. radiation 
emitted by electron before it is scattered by the proton, which
could provide measurement of form factors down to $Q^2=0.0001$ GeV$^2$ 
\cite{bernauer-phd}. Obviously, if this method could work, the 
proton radius would be severely constrained. 

\section{Effects of the nucleon radius variation on the symmetric 
  nuclear matter
  and the neutron star matter}
\label{sec:nuclear}
In the relativistic mean field (RMF) model the Lagrangian density 
of nucleons consists of four terms, i.e. the free nucleon, 
free meson, interaction between nucleons via meson exchange, and
meson self interaction terms. If we assume that electrons and muons are
point particles, whereas nucleons have structures with a radius 
$r_N$, then according to the RMF model 
the energy density of matter consisting of the nucleons and the leptons
is given by \cite{kouno,rischke}
\begin{eqnarray}
\epsilon&=&A (\epsilon_p^k + \epsilon_n^k) +\epsilon_e^k + \epsilon_{\mu}^k +\epsilon_M(\omega,\sigma,\rho)\nonumber\\
 &+& g_{\omega} \omega_0 (\rho_p+\rho_n)+\frac{1}{2} g_{\rho} b_0 (\rho_p-\rho_n),
\label{eq:eden} 
\end{eqnarray}
where 
$g_{\omega}$,  $g_{\sigma}$ and  $g_{\rho}$ are the couplings for
$\omega$, $\sigma$ and $\rho$ mesons, respectively, $\epsilon_M$ is the
total energy density of the meson, 
while $\sigma$, $\omega_0$ and 
$b_0$ are the $\sigma$, $\omega$ and $\rho$ fields, respectively. Furthermore, 
in Eq.~(\ref{eq:eden}) we have
\begin{equation}
\epsilon_i^k= \frac{2}{{(2 \pi)}^3} \int d^3\vec{k} \, 
{(k^2 +m^{*~2}_i)}^{1/2}\theta(k-k_F), ~~i=p,n,e,\mu ~,
\end{equation}
where for leptons the effective mass is 
$m^{*}_i$ =  $m_i$ and for nucleons  $m^{*}_i$= $m_i-g_{\sigma} \sigma$.

The nucleon and scalar densities read
\begin{eqnarray}
\rho_i= A \bar{\rho}_i ~,\\
\rho_{s,i}= A \bar{\rho}_{s,i} ~,
\end{eqnarray}
where  $\bar{\rho}_i$ and $\bar{\rho}_{s,i}$ are the $i$th nucleon and
scalar densities, assuming the nucleon is a point particle. The normalization
constant $A$ is given by
\begin{equation}
  \label{eq:norm}
  A=\frac{1}{1+V_p \bar{\rho}_p+V_n \bar{\rho}_n},
\end{equation}
with $V_p$ and  $V_n$ the proton and neutron volumes, respectively.
To simplify the present calculation we assume that 
\begin{eqnarray}
  \label{eq:radius-rmf}
  V_p=V_n\equiv V_N=\frac{4}{3}\pi r_N^3 ,
\end{eqnarray}
where $V_N$ and $r_N$ are the volume and radius of the
nucleon, respectively.

From Eq.~(\ref{eq:eden}) we can derive the matter pressure,
\begin{equation}
  P=\rho^2 \frac{d\varepsilon}{d\rho} ,
\end{equation}
with $\varepsilon=\epsilon/\rho$. Furthermore, the chemical potential
for the $i$th nucleon can be obtained from
\begin{equation}
\mu_i=E^*_{F~i}+V_i P_i'+ g_{\omega} \omega_0 +\alpha_i \frac{1}{2} g_{\rho} b_0 ,
\end{equation}
with $\alpha_i$ equals $+1$ ($-1$) for proton (neutron),
$E^*_{F,i}={(k_{F,i}^2 +m^{*2}_i)}^{1/2}$ and 
\begin{eqnarray}
 P_i'&=&\frac{1}{12 \pi^2} \left\{E^*_{F,i} k_{F,i}\left(E^{*2}_{F,i} -\frac{5}{2} m^{*2}_i\right)\right.\nonumber\\
     &+&\left. \frac{3}{2} m^{*4}_i \log\left(\frac{k_{F,i}+E^*_{F,i}}{m^*_i}\right)\right\} .
\end{eqnarray}

Different from the quark meson coupling (QMC) model \cite{panda,aguirre},
where the dependence of $r_N$ on the matter density can be directly 
obtained from the model, in the RMF approach the dependence cannot
be easily predicted. Therefore, in the present study we choose a 
phenomenological form for the nucleon radius, which is given by 
\begin{eqnarray}
  \label{eq:bag-radius}
  r_N(\rho)= {r_N(0)}{\left\{1+\beta\left(
        {\displaystyle \frac{\rho}{\rho_0}}\right)^2\right\}^{-2}} ,
\end{eqnarray}
where $\rho=\rho_p+\rho_n$, $\rho_0$ is the value of $\rho$
at saturation point, and $r_N(0)$ is the proton radius in vacuum
(zero density), determined from Eq.~(\ref{eq:rms}). At first glance, 
the choice seems to be trivial. However, it is actually selected 
to fulfill the causality
constraint. Furthermore, the formula given in Eq.~(\ref{eq:bag-radius})
is more convenient for the present purpose, rather than the exponential one, 
because in the framework of the presently used RMF model 
we found that the required
radius must slowly fall off as a function of density. Otherwise,
the predicted neutron star mass would violently overshoot the 
mass of the SRJ164-2230 pulsar, which is believed to be the 
heaviest observed neutron star \cite{demorest}. We
also observed that, for selected value of $\beta$, Eq.~(\ref{eq:bag-radius})
can be adjusted to mimic the result of the QMC models~\cite{aguirre}
in a certain range of density,
thus providing a good check of our result.

\begin{figure}[t]
  \begin{center}
    \leavevmode
    \epsfig{figure=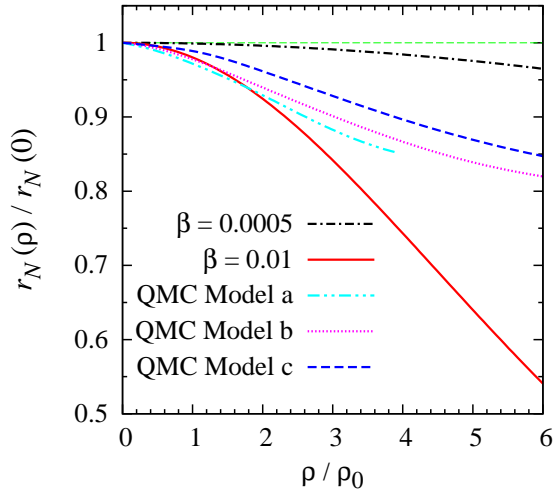,width=80mm}
    \caption{(Color online) Comparison between proton
      bag radii as a function of the ratio between nucleon and 
      nuclear saturation densities obtained from the QMC 
      models~\cite{aguirre} and Eq.~(\ref{eq:bag-radius}) with
      two different $\beta$ values.}
   \label{fig:rp} 
  \end{center}
\end{figure}

The ratio between $r_N(\rho)$ and $r_N(0)$ is exhibited in 
Fig.~\ref{fig:rp}, where we compare the results obtained from
two $\beta$ values, i.e. $\beta=0.0005$ and 0.01, with those
obtained from the QMC calculation using different values of bag
constants $B$ (see Ref.~\cite{aguirre} for explanation). It is
apparent from this figure that the difference between the 
calculated radii increases as the density increases. Nevertheless,
the distributions of the radii obtained from the QMC model
are still bounded within the difference of the two chosen $\beta$ values
in Eq.~(\ref{eq:bag-radius}), i.e. the solid and dash-dotted lines
in Fig.~\ref{fig:rp}.

For the RMF model we use the parameter set obtained by the 
IUFSU collaboration \cite{Fattoyev2010}.
In calculating the equation of state (EOS) of the neutron star matter
(NSM) we use the neutrality and $\beta$-stability conditions. They 
are required in calculating the Fermi momentum of each particle in the
neutron star core. 

The neutron star mass as a function of its radius
can be obtained by solving the Tolman-Oppenheimer-Volkoff (TOV)
equation with different particle densities in the neutron star core.
In order to describe the ''outer crust'' region we have used
the EOS given by  R\"uster {\it et al}.~\cite{Ruster2006}.
The EOS for the ''inner crust'' region is obtained from an 
extrapolation of the EOS of the core and ''outer crust'' 
by making use of the polytrophic
energy-pressure density approximation. The ''crust'' region is
described by using the RMF model with and without the EVE.

\begin{figure}[t]
  \begin{center}
    \leavevmode
    \epsfig{figure=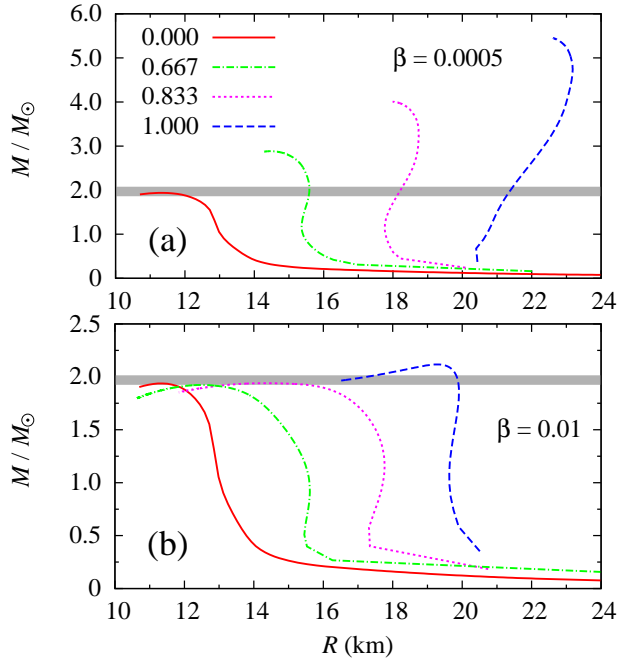,width=85mm}
    \caption{(Color online) The neutron star mass as a function
      of its radius for four different nucleon radius assumptions.
      Panels (a) and (b) show the calculation with 
      $\beta=0.0005$ and $\beta=0.01$ in Eq.~(\ref{eq:bag-radius}),
      respectively.
      The gray horizontal bands show the mass of the SRJ164-2230
      pulsar, which is believed to be the heaviest observed neutron
      star \cite{demorest}. 
      Notation of the curves in both panels 
      is given in Fig.~\ref{fig:eff-mass}. }
   \label{fig:mass-ns} 
  \end{center}
\end{figure}

Since in the nuclear and neutron star matter a direct 
comparison between model calculations and precise experimental 
data, as in the case of electron-proton scattering
in the previous section, is beyond
our imagination at present, in the followings we will 
not calculate the effect 
of averaging the nucleon radius on the possible observables.
Furthermore, as already obtained in the previous section,
the effect of this averaging process is a relatively tiny shift
from the original value.
Therefore, we believe that at this stage it is sufficient to 
investigate the effect on the conventional observables by using 
a $\pm 20\%$ variation of the nucleon original radius.
Note that this variation enters our calculation
through Eqs.~(\ref{eq:norm}) and (\ref{eq:radius-rmf}).

Figure \ref{fig:mass-ns} shows the sensitivity of  
the neuron star mass-radius relation to the variation of the
nucleon radius as well as to the dependence of the nucleon radius
on the matter density [$\beta$ in Eq.~(\ref{eq:bag-radius})],
in the framework of RMF models.
The radius of the neutron star depends on the value of
nucleon radius, whereas the maximum mass of the neutron star
is controlled by the dependence of the nucleon radius
on the matter density. Therefore, the heaviest observed neutron star mass 
PSRJ1614-2230 \cite{demorest} shown by the horizontal gray lines in
Fig.~\ref{fig:mass-ns} yields a significant suppression
of the nucleon radius at very high density. This result is
interesting, because it opens the possibility of hyperon
existence in a neutron star by using the hyperon vector
couplings obtained from SU(6) symmetry \cite{weissenborn}.

With regard to the radius of the canonical neutron star
($1.4M_\odot$), which is constrained between $10.4$ km and $12.9$ 
km \cite{andrew}, our present result should be carefully interpreted,
because in this calculation we have used the IUFSU parameter set, 
which was fitted to the finite nuclei data by 
assuming point particle approximation for the nucleon. Therefore,
the present result cannot be quantitatively 
compared with those obtained with other
constraints. Nevertheless, Fig.~\ref{fig:mass-ns} indicates that
the present result could become a stringent constraint
to the nucleon radius at high density, once a consistent
EOS with EVE were available.

\begin{figure}[h]
  \begin{center}
    \leavevmode
    \epsfig{figure=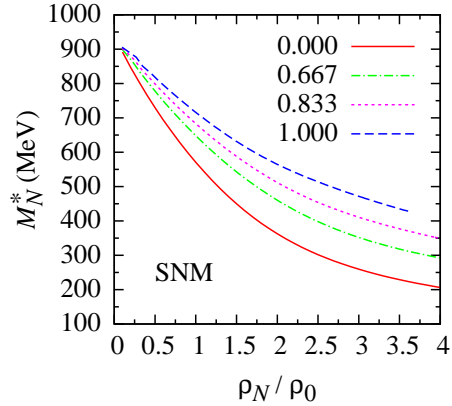,width=65mm}
    \caption{(Color online) Effective nucleon mass as a function
      of the ratio between nucleon and nuclear saturation densities.
      The results are obtained with different values of the 
      nucleon radius $r_N(0)$ as indicated in the figure (in the 
      unit of fm).}
   \label{fig:eff-mass} 
  \end{center}
\end{figure}

\begin{figure}[h]
  \begin{center}
    \leavevmode
    \epsfig{figure=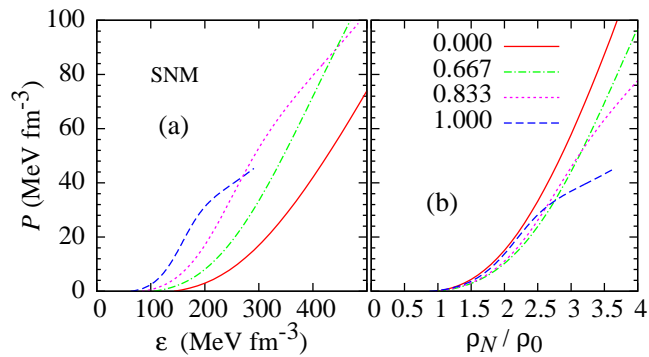,width=85mm}
    \caption{(Color online) Equation of states of the symmetric
    nuclear matter obtained from calculations with different
    values of the nucleon radius as a function of (a) energy
    and (b) density. Notation of the curves is as in 
    Fig.~\ref{fig:eff-mass}.}
   \label{fig:eos-snm} 
  \end{center}
\end{figure}

Since the result obtained by using $\beta=0.0005$ substantially 
overshoots the PSRJ1614-2230 constraint, as shown in 
Fig.~\ref{fig:mass-ns}, in the following discussion we will only use
$\beta=0.01$.
The result for the effective nucleon mass in the case of 
symmetric nuclear matter (SNM) is shown in Fig.~\ref{fig:eff-mass},
where we compare the calculated masses obtained by assuming
point particle approximation [$r_N(0)=0$] and finite nucleon 
radii [$r_N(0)\ne 0$]. 

From Fig.~\ref{fig:eff-mass} it is apparent that at high densities 
the nucleon effective
mass increases with increasing the nucleon radius.
In view of the instability against the particle-hole excitation at high
densities due to the density fluctuation \cite{anto}, 
a sufficiently large effective mass predicted
by the RMF model has an obvious advantage. 

\begin{figure}[t]
  \begin{center}
    \leavevmode
    \epsfig{figure=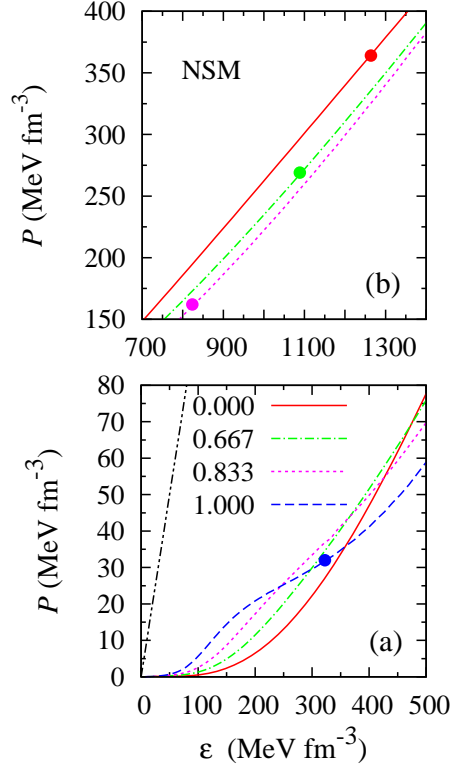,width=60mm}
    \caption{(Color online) As in Fig.~\ref{fig:eos-snm}a, but
      for the neutron star matter at (a) lower and (b) higher
      energies. The dash-dot-dotted line in panel (a) 
      shows the causality constraint, the solid circles
    indicate the center pressures and energy densities
  in the neutron star with maximum mass.}
   \label{fig:eos-nsm} 
  \end{center}
\end{figure}

\begin{figure}[t]
  \begin{center}
    \leavevmode
    \epsfig{figure=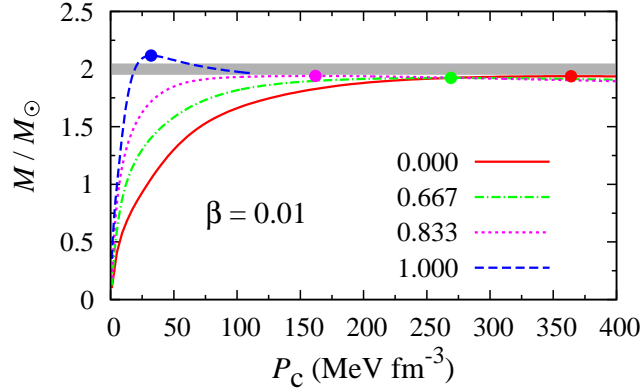,width=85mm}
    \caption{(Color online) Neutron star mass as a function of its
      center pressure for different nucleon radii obtained 
      with $\beta=0.01$.
    Solid circles indicate the maximum masses with the
    corresponding center pressures.}
   \label{fig:mass_pc} 
  \end{center}
\end{figure}

Figures \ref{fig:eos-snm} and \ref{fig:eos-nsm} exhibit the non-zero 
nucleon radius effect on the EOS of SNM and NSM, respectively. The 
solid circles in the NSM EOS of  Fig. \ref{fig:eos-nsm} indicate 
the positions of the NS center pressures 
and center energy densities of maximum mass. At moderate densities, which 
correspond to $\epsilon\lesssim 300$, we observe that the EOS becomes 
stiffer as the nucleon radius increases. Beyond this range, the EOS
tends to be softer. From Fig. \ref{fig:mass_pc}, it is obvious that the 
pressure of the NS center with maximum mass of $r_N(0)=1$ fm (blue dashed 
line) is approximately 32 MeV fm$^{-3}$, which corresponds to the energy density 
of about 320 MeV fm$^{-3}$. At this point the EOS obtained with 
$r_N(0)=1$ fm is 
stiffest compared to other cases [panel (a) of Fig. 10]. Since in 
obtaining the NS mass we should integrate the TOV equation using 
the corresponding EOS as input from the NS center pressure up to 
zero, information based solely on the soft EOS at high densities 
of $r_N(0)=1$ fm is insufficient for a complete understanding 
of the NS maximum mass. 
For other non zero nucleon radius
cases, the situation is similar. 

\begin{figure}[t]
  \begin{center}
    \leavevmode
    \epsfig{figure=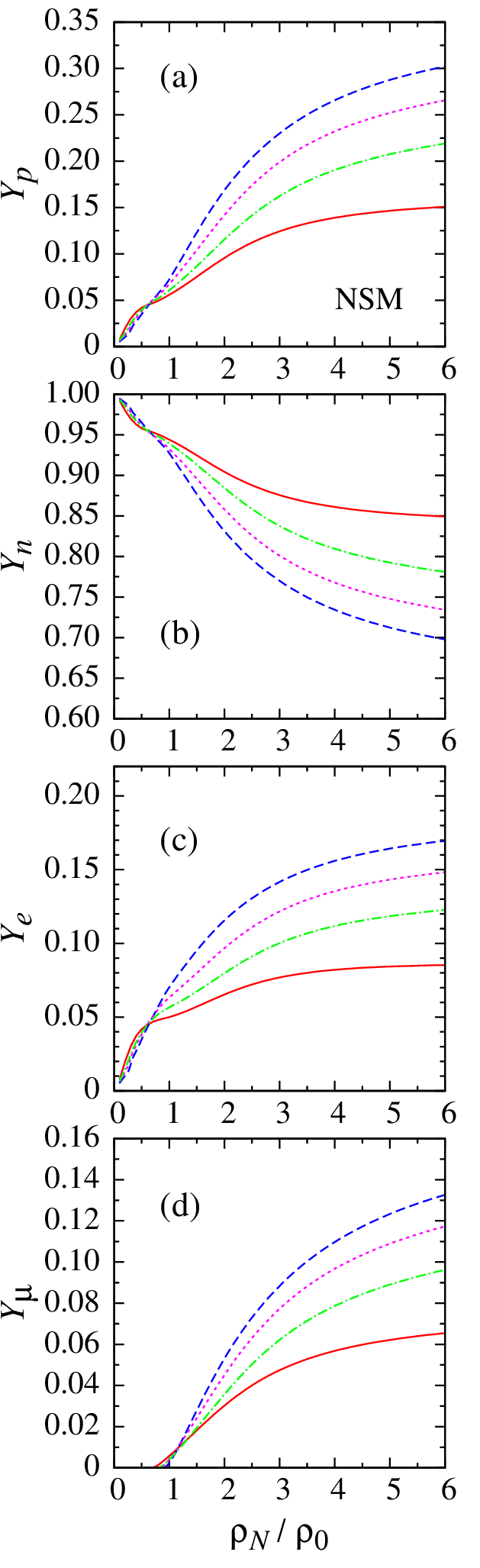,width=52mm}
    \caption{(Color online) Proton, neutron, electron, and muon
      fractions in the neutron star matter as a function 
      of the ratio between nucleon and nuclear saturation densities.
      Notation of the curves is as in Fig.~\ref{fig:eff-mass}.}
   \label{fig:frac-nsm} 
  \end{center}
\end{figure}

Obviously, decreasing the nucleon radius will decrease the pressure 
in the region of $\epsilon\le$ 300 MeV fm$^{-3}$. However, 
decreasing the nucleon radius will simultaneously increase 
the NS center pressure. Thus, additional 
contribution from the center pressure up to  300 MeV fm$^{-3}$ will only 
slightly increase the NS mass. 
This is due to the fact that the corresponding EOS at high 
densities are relatively soft. 
On the other hand, in the case of zero nucleon radius 
but using the same RMF parameter set, the contribution from  300 MeV 
fm$^{-3}$ up to zero is very small but the contribution from high 
densities is dominant because the corresponding EOS is stiffer than 
that of the nonzero nucleon radius. The very small contribution in 
the region $\le$ 300 MeV fm$^{-3}$ and the relatively large center 
pressure are typical for point-particle RMF models. In this case, 
the strong correlation between the NS maximum mass and the 
EOS stiffness at high densities is very obvious. Therefore, in the point 
particle case we only need to consider the EOS at high densities in  
investigating the NS maximum mass behavior. In addition, the 
high densities EOS become significantly stiffer as $\beta$ decreases,
As a consequence, decreasing the $\beta$ value will increase the 
maximum mass of NS. Thus, the increasing of the NS mass depends 
sensitively on the nucleon radius in free space. These 
phenomena explain the increase
of the predicted mass and radius of the neutron star with 
increasing the nucleon radius, as shown in Fig.~\ref{fig:mass-ns}.

In Fig.~\ref{fig:frac-nsm} we display the finite nucleon radius 
effect on the fraction of the matter constituents in the neutron star. It is
obvious from this figure that increasing the nucleon radius will
increase the number of existing charge particles at high density.
This result has a serious consequence on a number of neutron star 
properties, such as the neutron star stability, neutrino transport
in the neutron star, as well as the cooling process of a neutron
star. Unfortunately, a more detailed and quantitative analysis 
of the EVE on neutron star should wait for a more consistent 
EOS, which includes  the EVE in the calculation. 

\section{Summary and Conclusion}
\label{sec:conclusion}
We have investigated effects of the proton radius  variation 
on the extraction of the proton charge form factor. 
To achieve the best agreement with experimental data we have 
averaged a dipole form factor over the corresponding 
cut-off with an upper (lower) integration 
limit of $+21.5\%$ ($-21.5\%$) from its middle value. 
The extracted proton charge radius is found to be smaller than that obtained
using the traditional standard dipole fit, 
but is in good agreement with those obtained from 
a recent measurement of the Lamb shift in muonic hydrogen atom as 
well as  from the dispersion relation. The extracted proton magnetic
radius is smaller than the result of the dispersion relation, but in
agreement with the direct extraction by making use of the 
Friedrich-Walcher form factor. Nevertheless, since the magnetic form
factor is less accurate, the extraction of magnetic radius is 
also less reliable as compared to the result 
of the charge radius.

We have also investigated effects of the nucleon radius variation
on the SNM and the NSM. To this end, the nucleon
radius dependence on the matter density is described by a simple 
phenomenological form and four assumptions of the nucleon radius 
at zero density are considered in the calculation, i.e. 0 fm
(point particle approximation), 
0.833 fm (original radius), as well as 0.667 fm and 1.000 fm
($\pm 20\%$ modifications of the original radius). We found 
that the relation between the mass and radius of a neutron star 
is very sensitive to the radius of the nucleon. A similar 
result is also observed in the case of the effective nucleon
mass of the SNM  as well as the EOS of both NSM and 
SNM. However, a more quantitative conclusion  could be drawn
only after a more consistent EOS, with the EVE considered, had been 
available.

\section*{Acknowledgment}
This work has been supported in part by the University of Indonesia
and the Competence Grant of the Indonesian 
Ministry of Education and Culture.

\end{document}